\documentclass[journal,12pt,onecolumn]{IEEEtran}
\usepackage{bbm, enumitem, indentfirst}
\usepackage{amsmath, amssymb, amsthm}  
\usepackage{rotating, booktabs}  
\usepackage{caption, graphicx, array, subfigure, url, color}
\usepackage[ruled]{algorithm2e}
\usepackage{algpseudocode}
\usepackage{multirow}

\SetKwRepeat{Do}{do}{while}

\newcommand {\mymarginpar}[1]{\marginpar{#1}}
\renewcommand {\marginpar}[1]{}

\def\_{\rule{.3em}{.15ex}}      

\newcommand{\ls}[1]
   {\dimen0=\fontdimen6\the\font
    \lineskip=#1\dimen0
    \advance\lineskip.5\fontdimen5\the\font
    \advance\lineskip-\dimen0
    \lineskiplimit=.9\lineskip
    \baselineskip=\lineskip
    \advance\baselineskip\dimen0
    \normallineskip\lineskip
    \normallineskiplimit\lineskiplimit
    \normalbaselineskip\baselineskip
    \ignorespaces
   }

\newcommand\sgo{(G(V,E), p_{U,W}(\cdot,\cdot))}

\newtheorem{definition}{Definition}
\newtheorem{property}[definition]{Property}
\newtheorem{proposition}[definition]{Proposition}
\newtheorem{lemma}[definition]{Lemma}
\newtheorem{theorem}[definition]{Theorem}
\newtheorem{corollary}[definition]{Corollary}
\newtheorem{example}[definition]{Example}
\newtheorem{remark}[definition]{Remark}

\newcommand {\benum} {\begin{enumerate}}
\newcommand {\eenum} {\end{enumerate}}
\newcommand {\bdesc} {\begin{description}}
\newcommand {\edesc} {\end{description}}

\newcommand {\bfig}[2] {\begin{figure}
  \centering
  \includegraphics[width=#2]{#1}}
\newcommand {\brotatefig}[2] {\begin{figure}[htbp]
                        \centerline {
                         \epsfig{figure={#1},clip=,angle=-90,width={#2}}}}

\newcommand {\bfigfirst}[2] {\begin{figure}[h]
                        \centerline {
                        \setlength{\epsfxsize}{#2}
                        \epsffile{#1}}}
\newcommand {\efig}[2]{ \caption{#2}
                        \label{fig:#1}
                        \end{figure}
                        \mymarginpar{fig:#1}}
\newcommand {\erotatefig}[2]{ \caption{#2}
                        \label{fig:#1}
                        \end{figure}
                        \mymarginpar{fig:#1}}
\newcommand {\rfig}[1]{Figure \ref{fig:#1}}

\newcommand {\btab}[1]{
                       \begin{table}
                       \centering
                       \begin{tabular}{#1}}
\newcommand {\etab}[3] {
                       \end{tabular}
                       \caption[#3]{#2}
                       \label{tab:#1}
                       \end{table}
                       \mymarginpar{tab:#1}
                       \vspace{.1in}}
\newcommand {\rtab}[1]{Table \ref{tab:#1}}

\newcommand {\btabular}[1]{\begin{center}
                       \begin{tabular}{#1}}
\newcommand {\etabular}{\end{tabular}
                       \end{center}}

\newcommand {\bdefin}[1]{\begin{definition}
                      \mymarginpar{def:#1}
                      \label{def:#1} }
\newcommand {\edefin}       {\end{definition}}

\newcommand {\bpro}[1]{\begin{property}
                      \mymarginpar{pro:#1}
                      \label{pro:#1} }
\newcommand {\epro}   {\end{property}}

\newcommand {\bprop}[1]{\begin{proposition}
                      \mymarginpar{prop:#1}
                      \label{prop:#1} }
\newcommand {\eprop}       {\end{proposition}}

\newcommand {\blem}[1]{\begin{lemma}
                      \mymarginpar{lem:#1}
                      \label{lem:#1} }
\newcommand {\elem}   {\end{lemma}}

\newcommand {\bthe}[1]{\begin{theorem}
                      \mymarginpar{the:#1}
                      \label{the:#1} }
\newcommand {\ethe}   {\end{theorem}}

\newcommand {\bcor}[1]{\begin{corollary}
                      \mymarginpar{cor:#1}
                      \label{cor:#1} }
\newcommand {\ecor}   {\end{corollary}}

\newcommand {\bax}[1]{\begin{axiom}
                      \mymarginpar{ax:#1}
                      \label{ax:#1} }
\newcommand {\eax}       {\vspace{-.1in} \end{axiom}}

\newcommand {\bex}[2]{\vspace{.1in}
                      \begin{example}
                      \mymarginpar{ex:#1}
                       {\bf #2}
                      \label{ex:#1} }
\newcommand {\eex}       {\end{example} \vspace{.3cm} }
\newcommand {\rex}[1]{Example \ref{ex:#1}}

\newcommand {\brem}[1]{\begin{remark}
                      \mymarginpar{rem:#1}
                      \label{rem:#1} \em }
\newcommand {\erem}   {\end{remark}}

\newcommand {\beq}[1]{\mymarginpar{eq:#1}
                      \begin{equation}
                      \label{eq:#1} }

\newcommand {\beqno}[1]{\mymarginpar{eq:#1}
                      \begin{eqnarray}
                      \nonumber}

\newcommand {\eeq}       {\end{equation}}
\newcommand {\eeqno}       { && \end{eqnarray}}
\newcommand {\req}[1]{(\ref{eq:#1})}

\newcommand {\bear}[1]{\mymarginpar{eq:#1}
                       \begin{eqnarray}
                       \label{eq:#1} }

\newcommand {\bearno}[1]{\mymarginpar{eq:#1}
                       \begin{eqnarray}
                       \nonumber}

\newcommand {\eear}{\end{eqnarray}}
\newcommand {\eearno}{\end{eqnarray}}

\newcommand {\bsel}{\left \{ \begin{array}{cl}}
\newcommand {\esel}{\end{array} \right.}

\newcommand {\bmat}[1]{\left [ \begin{array}{#1}}
\newcommand {\emat}{\end{array} \right ]}

\def\R{I\kern-0.30em R}
\def\N{I\kern-0.30em N}
\def\P{I\kern-0.30em P}
\def\pr{{\bf\sf P}}

\IEEEoverridecommandlockouts

\begin{document}

\title{Exponentially Twisted Sampling: a Unified Approach for Centrality Analysis in Attributed Networks}

\date{July, 2017}

\author{Cheng-Hsun~Chang and~Cheng-Shang~Chang

\thanks{
C.-H. Chang and C.-S. Chang are with the Institute of Communications Engineering,
National Tsing Hua University, Hsinchu 30013, Taiwan, R.O.C.
Email: s104064525@m104.nthu.edu.tw; cschang@ee.nthu.edu.tw.}}

\maketitle

\begin{abstract}

In our recent works, we developed a probabilistic framework for structural analysis in {\em undirected networks} and {\em directed networks}. The key idea of that framework is to sample a network by a {\em symmetric} and {\em asymmetric} bivariate distribution and then use that bivariate distribution to formerly defining various notions, including centrality, relative centrality, community, and modularity. The main objective of this paper is to extend the probabilistic definition to {\em attributed} networks, where sampling bivariate distributions by exponentially twisted sampling. Our main finding is that we find a way to deal with the sampling of the attributed network including signed network. By using the sampling method, we define the various centralities in attributed networks. The influence centralities and trust centralities correctly show that how to identify centralities in signed network. The advertisement-specific influence centralities also perfectly define centralities when the attributed networks that have node attribute. Experimental results on real-world dataset demonstrate the different centralities with changing the temperature. Further experiments are conducted to gain a deeper understanding of the importance of the temperature.

\end{abstract}


\IEEEpeerreviewmaketitle

\section{Introduction}
\IEEEPARstart{I}{n} social network analysis, centrality \cite{freeman1977set,newman2010networks,freeman1978centrality} have been widely used for ranking the important nodes. To obtain a viewpoint of a network, one typical method is to "sample" the network, e.g., edge sampling, random walks, diffusion \cite{mucha2010community}, or random gossiping \cite{boyd2005gossip}. Mathematically, each sampling method renders a (probability) measure for a network that enables us to carry out further analysis. As in the probabilistic framework in \cite{chang2011general}, we model a network as a graph $G$ and “sample” the graph to generate a bivariate distribution $p(\cdot,\cdot)$ for a pair of two nodes.  In this paper, we only consider the case that the bivariate distribution is symmetric. Under this assumption, the two marginal distributions of the bivariate distribution are the same. As such, the marginal distribution can be used for defining the centrality. Then, we can use centrality to rank nodes. Ranking nodes in the unsigned network have existed a long time such as PageRank \cite{page1999pagerank}. There are many papers in the development of new centralities has made great progress in a static network \cite{latora2007measure,estrada2005subgraph,mohammadinejad2016employing}. Moreover, they extend network properties to the dynamic network \cite{kim2012temporal,taylor2017eigenvector}. Also, many researchers have proposed heuristics on the standard methods to make them computable in signed networks \cite{jung2016personalized} and attributed network \cite{weng2010twitterrank}, but there is no unified approach for centrality. In this paper, we propose "exponentially twisted sampling," the exponential change of measure to get the bivariate distribution $p(\cdot,\cdot)$, to sample the graph $G$. We show that our centrality measure digs out various centrality, including the influence centralities, the trust centralities, the advertisement-specific influence centralities, and homophily centrality.

The main objective of this paper is to extend the centrality in \cite{chang2011general,chang2015relative,chang2016probabilistic} to {\em attributed} networks which contain edge attribute and node attribute. Our main finding is that we can find the {\em path measure} $f$ to sample the path $r$. For this, we show that one can construct another sampled graph with a {\em exponentially twisted sampling}, and then the path measure $f$ can be represented as a function of attributes of the nodes and edges along the path. In this paper, we also address a method for sampling an attributed network with a bivariate distribution and a path measure.

\section{Sampled graph}

\subsection{Review of the probabilistic framework of sampled graphs}

In \cite{chang2015relative,chang2016probabilistic}, a probabilistic framework for structural  analysis in undirected/directed networks was proposed. The main idea in that framework is to sample  a network by randomly selecting a path in the network. A network with a path sampling distribution is then called a {\em sampled graph} in \cite{chang2015relative,chang2016probabilistic} that can in turn be used for structural analysis of the network, including centrality and community.
Specifically, suppose a network is modeled by a graph $G(V,E)$, where $V$ denotes the set of vertices (nodes) in the graph and $E$ denotes the set of edges (links) in the graph. Let $n=|V|$ be the number of vertices in the graph and index the $n$ vertices from $1,2,\ldots, n$. Also, let $A=(a_{i,j})$ be the $n \times n$ adjacency matrix of the graph, i.e.,
\bearno
aa_{i,j}
=  \left\{\begin{array}{ll}
                 1, & \mbox{if there is an edge from  vertex {\em i} to vertex {\em j}}, \\
                 0, & $otherwise$.
                \end{array} \right.
\eearno
Let $R_{u,w}$ be the set of (directed) paths from $u$ to $w$ and $R=\cup_{u,w \in V} R_{u,w}$ be the set of paths in the graph $G(V,E)$. According to a probability mass function $p(\cdot)$, called the {\em path sampling distribution}, 
a path $r \in R$ is selected at random with probability $p(r)$.
In \cite{chang2015relative,chang2016probabilistic}, there are many  methods for sampling a graph with a randomly selected path.
Here we introduce the following three commonly used approaches: (i) sampling by uniformly selecting a directed edge, (ii) sampling by a Markov chain, and (iii) sampling by a random walk on an undirected network with path length 1 or 2.

\bex{uniform}{\bf (Sampling by uniformly selecting a directed edge)}
Given a directed graph $G=(V,E)$ with the adjacency matrix $A=(a_{i,j})$, one only sample directed paths with length 1 and this is done uniformly among all the directed edges. Specifically, sampling by uniformly selecting a directed edge has the following probability mass function:
\bearno
pp(r)
=  \left\{\begin{array}{ll}
                 1/m, & \mbox{if $r$ is an edge from  vertex {\em i} to vertex {\em j}}, \\
                 0, & $otherwise$.
                \end{array} \right.,
\eearno
where $m=|E|$ is the total number of directed edges in the graph.
\eex

\bex{MC}{\bf (Sampling by a Markov chain)}
Given a directed graph $G=(V,E)$ with the adjacency matrix $A=(a_{i,j})$, consider an ergodic Markov chain on this graph.
Let $p_{u,w}$ be the transition probability from node $u$ to node $w$ and $\pi_u$ be the steady state probability of node $u$.
In particular, for PageRank \cite{brin2012reprint} with the web surfing probability $\lambda$, the transition probability of the corresponding Markov chain is
\beq{Page0000}
p_{u,w}= (1-\lambda)\frac{1}{n} + \lambda \frac{a_{u,w}}{k_u^{out}},
\eeq
where $k_u^{out}$ is the out-degree of node $u$. Its steady state probabilities (with $\sum_{u=1}^n \pi_u=1$) can be obtained from solving the following system of equations:
\beq{Page1111}
\pi_u =(1-\lambda)\frac{1}{n} + \lambda \sum_{w=1}^n \frac{a_{wu}}{k_w^{out}} \pi_w ,\quad \mbox{for all}\;u=1,2 , \ldots, n.
\eeq
For a path $r$ that traverses a sequence of nodes $\{u_1, u_2,\ldots, u_{k-1}, u_k\}$ in a directed network, we have from the Makrov property that
\beq{MC1111}
p(r)=\pi_{u_1}\cdot p_{u_1,u_2} \cdot \ldots \cdot p_{u_{k-1}, u_k}.
\eeq
Consider the reverse path of $r$, denoted by $Rev(r)$, that traverses a sequence of nodes $\{u_k, u_{k-1},\ldots, u_{2}, u_1\}$.
If the Markov chain is reversible, then it follows from the detailed balance equation that
\beq{MC2222}
p(r)=p(Rev(r)).
\eeq
It is known that the corresponding Markov chain for PageRank is in general not reversible. However, the corresponding Markov chain for a random walk on an undirected graph is reversible. We will discuss this further in the next example.
\eex

\bex{path2b}{\bf (Sampling by a random walk on an undirected network with path length 1 or 2)}
For an undirected graph $G(V,E)$, let $m=|E|$ be the total number of edges and $k_v$ be the degree of node $v$, $v=1,2,\ldots, n$. A path $r$ with length 1 can be represented by the two nodes  $\{u_1,u_2\}$ it traverses. Similarly, a path with length 2 can be represented by the three nodes $\{u_1,u_2,u_3\}$ it traverses.
A random walk with path length not greater than 2 can be generated by the following two steps: (i) with the probability $k_v/2m$, an initial node $v$ is chosen, (ii) with probability $\beta_i$, $i=1,2$, a walk with length $i$ is chosen. As such, we have
\bear{path2222}
p(r)=\left\{\begin{array}{ll}
                 {{\beta_1} \over {2m}}a_{u_1,u_2}, & \mbox{if}\; r=\{u_1,u_2\}, \\
                 \\
                {{\beta_2} \over {2m}} \frac{a_{u_1, u_2} a_{u_2,u_3}}{k_{u_2}}, & \mbox{if}\; r=\{u_1,u_2,u_3\}.
                \end{array} \right.
\eear
where $ \beta_1 +\beta_2=1$ and $\beta_i \ge 0$, $i=1,2$.
For an undirected network, we have $a_{i,j}=a_{j,i}$ for all $i,j=1,2,\ldots, n$. Thus, in view of \req{path2222}, we also have
\beq{path2222r}
p(r)=p(Rev(r))
\eeq
in this example.
\eex

Let $U$ (resp. $W$) be the starting (resp. ending) node of a randomly select path by using the path sampling distribution $p(\cdot)$.
Then the bivariate distribution
\beq{frame1111}
p_{U,W}(u,w)=\pr(U=u,W=w)=\sum_{r\in R_{u,w}} p(r)
\eeq
is the probability that the ordered pair of two nodes $(U,W)$ is selected.   As such, $p_{U,W}(u,w)$ can be viewed as a similarity measure from node $u$ to node $w$  and this leads to the definition of a sampled graph in \cite{chang2015relative,chang2016probabilistic}.
In general, the bivariate distribution  in \req{frame1111} is not {\em symmetric} by using the path sampling distributions in \rex{uniform} and \rex{MC}.
However, if $p(r)=p(Rev(r))$ for any path $r$, then the bivariate distribution  in \req{frame1111} is clearly symmetric.


\bdefin{sampled}{\bf (Sampled graph \cite{chang2015relative,chang2016probabilistic})} A  graph $G(V,E)$ that is sampled by randomly selecting an ordered pair of two nodes $(U,W)$ according to a specific bivariate distribution $p_{U,W}(\cdot,\cdot)$  in \req{frame1111} is called a {\em sampled graph} and it is denoted by the two-tuple $\sgo$.
\edefin

Let $p_U(u)$ (resp. $p_W(w)$) be the marginal distribution of the random variable $U$ (resp. $W$), i.e.,
\beq{frame2222}
p_U(u)=\pr(U=u)= \sum_{w=1}^n p_{U,W}(u,w),
\eeq
and
\beq{frame3333}
p_W(w)= \pr(W=w)=\sum_{u=1}^n p_{U,W}(u,w).
\eeq
Then $p_U(u)$ is the probability that node $u$ is selected as a starting node of a path and it can be viewed as an out-centrality of $u$. On the other hand, $p_W(w)$ is the probability that node $w$ is selected as a ending node of a path and it can be viewed as an in-centrality of $w$.
The in-centrality and the out-centrality  are in general not the same.
Clearly, if the bivariate distribution $p_{U,W}(\cdot,\cdot)$ is symmetric, then the in-centrality and the out-centrality are the same.
A recent advance in \cite{chang2016probabilistic} shows that one does not need a symmetric bivariate distribution to ensure the equality between the in-centrality and the out-centrality. In particular,
for the Markov chain sampling methods in \rex{MC}, one still has $p_U(u)=p_W(u)$ and the in-centrality and the out-centrality are the same. In that case, we will simply refer $P_U(u)$ as the centrality of node $u$.

The framework of sampled graphs in \cite{chang2015relative,chang2016probabilistic} can further be used to define the notions of {\em community} and {\em modularity}. 
There are two different types of communities for directed networks that are commonly addressed in the literature (see e.g., \cite{rosvall2007maps,lambiotte2014random}): structural communities (based on the densities of edges) and flow communities (based on the flow of probabilities). 
For (probabilistic) flow communities, they can be detected by using Markov chains in \rex{MC}. On the other hand, structural communities can be detected by using uniform edge sampling in \rex{uniform} to produce communities that are densely connected inside and sparsely connected outside.
These two notions of {\em community} and {\em modularity}  will not be pursued further in this paper.

\subsection{Exponentially twisted sampling}

Now we generalize the probabilistic framework in \cite{chang2015relative,chang2016probabilistic} to attributed networks.
An attributed network is a generalization of a graph $G(V,E)$ by assigning each node $u \in V$ an attribute  $h_V(u)$ and each edge $e \in E$ an attribute $h_E(e)$. As such,  an attributed network can be represented as $G(V,E,h_V(\cdot),h_E(\cdot))$, where $h_V(\cdot)$ and $h_E(\cdot)$ are called the node attribute function and the edge attribute function, respectively.

For a path $r$ that traverses a sequence of nodes $\{u_1, u_2,\ldots, u_{k-1}, u_k\}$ in a attributed network, we can define a path measure
$f(r)$ as a function of the attributes of the nodes and edges along the path $r$, i.e.,
$$\{h_V(u_1),\ldots, h_V(u_k), h_E(u_1,u_2),\ldots, h_E(u_{k-1},u_k)\}.$$
In this paper, we assume that a path measure $f(\cdot)$ is a mapping from the set of paths $R$ to a $p$-dimensional real-valued vector in ${\cal R}^p$, i.e.,
$$f(r)=(f_1(r), f_2(r), \ldots, f_p(r)),$$
where $f_i(\cdot)$, $i=1,2,\ldots, p$, are real-valued functions.

Now suppose that we have already had a sampled graph $(G(V, E), p_0(\cdot))$ that uses the probability mass function $p_0(r)$ to sample a path $r$ in $G(V,E)$. The question is how the sampling distribution should be changed so that the average path measure is equal to a specified vector $\bar f$. In other words, what is the most likely sampling distribution $p(\cdot)$ that leads to the average path measure $\bar f$ given that the original sampling distribution is $p_0(\cdot)$? For this, we introduce the
 Kullback-Leibler distance between two probability mass functions $p(\cdot)$ and $p_0(\cdot)$:
\beq{sign3333}
D(p \Vert p_0)=\sum_{r \in R}p(r) \log (\frac{p(r)}{p_0(r)}).
\eeq
The Kullback-Leibler distance is known to be nonnegative and it is zero if and only if $p(\cdot)=p_0(\cdot)$ (see e.g., \cite{cover2012elements}). Also, according to the Sanov theorem (see e.g., the books \cite{cover2012elements,chang2012performance}), the larger the Kullback-Leibler distance is, the more unlikely for
  $p_0(\cdot)$ to behave like $p(\cdot)$. Thus, to address the question, we consider the following constrained  minimization problem:
\bear{sign4444}
&\min \quad &D(p \Vert p_0)\nonumber \\
&s.t. \quad &\sum_{r\in R}p(r)=1, \nonumber \\
&&\sum_{r\in R}f(r)p(r)=\overline{f}.
\eear
\\
The first constraint states that the total probability  must be equal to $1$. The second constraint states that the average path measure must be equal to $\bar f$ with the new sampling distribution $p(\cdot)$.
The above minimization problem can be solved by using Lagrange's multipliers $\alpha \in {\cal R}$ and $\theta \in {\cal R}^p$ as follows:
\bear{sign5555}
I=D(p \Vert p_0)+\alpha(1-\sum_{r\in R}p(r))+\theta \cdot (\overline{f}-\sum_{r\in R}f(r)p(r)).
\eear
Taking the partial derivative with respect to $p(r)$ yields
\beq{sign6666}
\frac{\partial I}{\partial p(r)}= \log p(r)+1-\log p_0(r)-\alpha-\theta \cdot f(r)=0.
\eeq

Thus,
\beq{sign7777}
p(r)=\exp({\alpha-1})*\exp({\theta \cdot f(r)})*p_0(r).
\eeq
Since $\sum_{r\in R}p(r)=1$,
it then follows that
\beq{sign8888}
p(r)=C*\exp(\theta \cdot f(r))*p_0(r),
\eeq
where
\beq{sign8888b}
C=\frac{1}{\sum_{r \in R}\exp(\theta \cdot f(r))*p_0(r)}
\eeq
is the normalization constant. The new sampling distribution in \req{sign8888} is known as the exponentially twisted distribution  in the literature (see e.g., \cite{juneja2006rare}).

To solve  $\theta$, we  let
$$
F=\log(1/C).
$$
The quantity $F$ is called the free energy as it is analogous to the free energy in statistical mechanic \cite{newman2010networks}.
It is easy to see that for $i=1,2,\ldots
,p$ that
\beq{partialdifferentiation}
\frac{\partial F}{\partial \theta_i}=\sum_{r\in R}f_i(r)p(r)=\overline{f}_i .
\eeq
These $p$ equations can then be used to solve $\theta_i$, $i=1,2,\ldots, p$.

Once we have the sampling distribution in \req{sign8888}, we can define a bivariate distribution $p_{U,W}(u,w)$ as in \req{frame1111}.
Specifically, let $U$ (resp. $W$) be the starting (resp. ending) node of a randomly select path according to the sampling distribution in \req{sign8888}.
Then the bivariate distribution
\beq{frame1111b}
p_{U,W}(u,w)=\pr(U=u,W=w)=\sum_{r\in R_{u,w}} p(r)
\eeq
is the probability that the ordered pair of two nodes $(U,W)$ is selected and it can be viewed as a similarity measure from node $u$ to node $w$.
Analogous to the discussion of a sampled graph in the previous section, 
the  marginal distribution of the random variable $U$ (resp. $W$), i.e., $p_U(u)$ (resp. $p_W(w)$), can be
viewed as an out-centrality of $u$ (resp. in-centrality of $w$). 

\section{Centralities}

\subsection{Centralities in signed networks}
In this section, we consider a special class of attributed networks, called {\em signed networks}.
A signed  network $G=(V,E,h_E(\cdot))$ is an attributed network with an edge attribute function $h_E(\cdot)$ that maps
every undirected edge in $E$ to the two signs $\{+,-\}$. In this paper, we represent the positive (resp. negative) sign  by 1 (resp. -1). An edge $(u,w)$ mapped with the $+$ sign  is called a {\em positive} edge, and it is generally used for indicating the {\em friendship} between the two nodes $u$ and $w$. On the other hand, an edge mapped with the $-$ sign  is called a {\em negative} edge. A negative edge $(u,w)$ indicates that $u$ and $w$ are {\em enemies}.

\subsubsection{Influence centralities}

Such a problem was previously studied in \cite{jung2016personalized} for ranking nodes in signed networks. In SRWR \cite{jung2016personalized}, a signed random surfer was used. They use the balance theory to imply the signed random walks. In our paper, it is our special case because it is as same as our how to sampling the path. In \cite{jung2016personalized}, it divides the equation into positive surfer $r^+_u$ and negative surfer $r^-_u$. However, we can consider positive and negative situations in a comprehensive way. 

One interesting question for signed networks is how the nodes in signed networks are ranked.
Our idea for this  is to 
 use opinion dynamics. If $u$ and $w$ are connected by a positive (resp. negative) edge, then it is very likely that $u$ will have a positive (resp. negative) influence on $w$ and vice versa. As such, if we start from a node $u$ with a positive opinion on a certain topic, then a neighbor of node $u$ connected by a positive (resp. negative) edge  will  tend to have the same (resp. the opposite) opinion as node $u$ has. Now we can let the opinion propagate through the entire network (via a certain opinion dynamic) and count the (expected) number of nodes that have the same opinion as node $u$ has. If such a number is large, then it seems reasonable to say that node $u$ has a large positive influence on the other nodes in the network. In other words, a node $u$ has a large positive influence if there is a high probability that the other end of a randomly selected path has the same opinion as node $u$.
 This then leads us to define the notion of {\em influence centralities} for ranking nodes in signed networks.

The above argument is based on the general belief that ``a friend of my friend is likely to be my friend'' and ``an enemy of my enemy can be my friend'' in \cite{{newman2010networks}}. As such, for a path $r$  that traverses a sequence of nodes $\{u_1, u_2,\ldots, u_{k-1}, u_k\}$ in a signed network, we define the following path measure as the product of the edge signs along the path, i.e.,
\beq{path1111}
f(r)=\prod_{(u_i,u_{i+1})\in r}h_E({u_i,u_{i+1}}).
\eeq
Note that $f(r)$ is either 1 or -1 as the edge attribute function $h_E(\cdot)$ that maps
every undirected edge in $E$ to  $\{1,-1\}$.
In addition to the path measure, we also need to specify the probability that the path $r$ is selected, i.e., $p_0(r)$ in \req{sign8888}.
For this, we consider a random walk with path length 1 or 2 in \rex{path2b}.
It then follows from \req{sign8888}, \req{path1111} and \req{path2222} that
\bearno
pp(r)=\left\{\begin{array}{ll}
                 C \cdot e^{\theta h_E(u_1,u_2)} \cdot {{\beta_1} \over {2m}}a_{u_1,u_2}, & \\ \quad\quad \mbox{if}\; r=\{u_1,u_2\}, & \\
                 \\
                C \cdot e^{\theta \cdot h_E(u_1,u_2) \cdot h_E(u_2,u_3)} \cdot {{\beta_2} \over {2m}} \frac{a_{u_1, u_2} a_{u_2,u_3}}{k_{u_2}}, & \\ \quad\quad \mbox{if}\; r=\{u_1,u_2,u_3\}. &
                \end{array} \right.
\eearno
\beq{path3333}
\eeq
The constant $C$ in \req{path3333} is the normalization constant. Summing all the paths from $u$ to $w$ yields the bivariate distribution
\bear{path4444}
p_{U,W}(u,w)=C \Big [ e^{\theta h_E(u,w)} \cdot {{\beta_1} \over {2m}}a_{u,w} +\sum_{u_2 \in V} e^{\theta \cdot h_E(u,u_2) \cdot h_E(u_2,w)} \cdot {{\beta_2} \over {2m}} \frac{a_{u, u_2} a_{u_2,w}}{k_{u_2}}\Big ].
\eear
The marginal distribution of the bivariate distribution, denoted by $P_U(u)$, is the {\em influence centrality} of node $u$ (with respect to the temperature $\theta$).

If we  only select paths with length 1, i.e., $\beta_2=0$ in \req{path4444}, then there is a closed-form expression for the influence centrality. For this, we first compute the normalization constant $C$ by summing over $u$ and $w$ in \req{path4444} and this yields
\beq{path5555}
C= \frac{m}{m^+ e^\theta+m^-e^{-\theta}},
\eeq
where $m^+$ (resp. $m^-$) is the total number of positive (resp. negative) edges in the graph.
Thus, for $\beta_2=0$,
\bear{path6666}
p_U(u)&=&\sum_{w \in V} p_{U,W}(u,w) \nonumber \\
&=&\frac{(k_u^+ e^\theta+k_u^-e^{-\theta})}{2(m^+ e^\theta+m^-e^{-\theta})},
\eear
where $k_u^+$ (resp. $k_u^-$) is the number of positive (resp. negative) edges of node $u$.

Now suppose  we require  the average path measure $\bar f$ to be equal to some fixed constant $-1<\gamma<1$, then
we have from \req{partialdifferentiation} that
\bear{path7777}
\gamma&=&\bar f =\frac{\partial F}{\partial \theta}
\nonumber \\
&=&\frac{m^+\exp(\theta)-m^-\exp(-\theta)}{m^+\exp(\theta)+m^-\exp(-\theta)},
\eear
where $F=\log(1/C)$ with $C$ in \req{path5555} is the free energy.
This then leads to
\beq{random9999}
\theta=\ln(\sqrt[2]{\frac{m^-(1+\gamma)}{m^+(1-\gamma)}}).
\eeq

\subsubsection{Trust centralities}

As discussed in the previous section, the influence centralities are based on the general belief that ``an an enemy of my enemy can be my friend.'' Such a statement might be valid for modelling opinion dynamics. However, it is not suitable for modelling {\em trust}. In addition to the interpretation of a signed edge as the friend/enemy relationship, another commonly used interpretation is the trusted/untrusted link.
A path $r$ that traverses a sequence of nodes $\{u_1, u_2,\ldots, u_{k-1}, u_k\}$ can be {\em trusted}  if every edge is a trusted link so that there exists a chain of trust. In view of this, we consider another
path measure as the minimum of the edge signs along the path, i.e.,
\beq{path1111t}
f(r)=\min_{(u_i,u_{i+1})\in r}h_E({u_i,u_{i+1}}).
\eeq
If we use the random walk with path length 1 or 2 to sample a path $r$ in $G(V,E)$ as in \req{path2222}, then
the sampling distribution for the signed network with the path measure in \req{path1111t} can be written as follows:
\bearno
pp(r)=\left\{\begin{array}{ll}
                 C \cdot e^{\theta h_E(u_1,u_2)} \cdot {{\beta_1} \over {2m}}a_{u_1,u_2}, & \\ \quad\quad \mbox{if}\; r=\{u_1,u_2\}, & \\
                 \\
                C \cdot e^{\theta \cdot \min[h_E(u_1,u_2) , h_E(u_2,u_))]} \cdot {{\beta_2} \over {2m}} \frac{a_{u_1, u_2} a_{u_2,u_3}}{k_{u_2}}, & \\ \quad\quad \mbox{if}\; r=\{u_1,u_2,u_3\}. &
                \end{array} \right.
\eearno
\beq{path3333t}
\eeq
Moreover, we have the following bivariate distribution
\bear{path4444t}
p_{U,W}(u,w)=C \Big [ e^{\theta h_E(u,w)} \cdot {{\beta_1} \over {2m}}a_{u,w} +\sum_{u_2 \in V} e^{\theta \cdot \min[h_E(u,u_2) , h_E(u_2,w)]} \cdot {{\beta_2} \over {2m}} \frac{a_{u, u_2} a_{u_2,w}}{k_{u_2}}\Big ],\nonumber\\
\eear
where $C$ is the normalization constant.
The marginal distribution of the bivariate distribution, denoted by $P_U(u)$, is the {\em trust centrality} of node $u$ (with respect to the temperature $\theta$).
We note that if  we  only select paths with length 1, i.e., $\beta_2=0$,  then the trust centrality is the same as the influence  centrality in \req{path6666}.

\subsection{Advertisement-specific influence centralities in networks with node attributes}

In this section, we consider another class of attributed networks that have node attributes.
For a graph $G(V,E)$ with the node attribute function $h_V(u)$ that maps every  node $u$ to a vector in ${\cal R}^p$
\beq{topic1111}
(h_{V,1}(u), h_{V,2}(u), \ldots, h_{V,p}(u)).
\eeq
One intuitive way to interpret  such an attributed network is to view the graph $G(V,E)$ as a social network with $n$ users and the attribute vector in \req{topic1111} as the scores of  user $u$
 on various topics. Now suppose an advertisement $z$ can be represented by a vector of scores $(z_1,z_2, \ldots, z_p)$ with $z_i$ being the score of the $i^{th}$ topic. Then we would like to find out who are the most influential users in the network to pass on the advertisement $z$.
Such a problem was previously studied in \cite{weng2010twitterrank} for ranking nodes in Twitter.  In TwitterRank \cite{weng2010twitterrank}, a two-step approach was used. First, a topic-specific ranking is obtained for each topic by using a random surfer model similar to that in PageRank \cite{brin2012reprint}.
The second step is then to take the weighted average over these topic-specific rankings. Specifically, suppose that $RT_i(u)$ is the ranking for topic $i$ and user $u$. TwitterRank for  advertisement $z$ and user $u$ is then defined as the following weighted average:
\beq{topic2222}
\sum_{i=1}^p z_i RT_i(u).
\eeq
One  flaw for such a two-step approach is that it neglects the fact that the propagation of a specific advertisement through a user depends on how much a user ``likes'' the advertisement. To model how much 
a user ``likes'' an advertisement, we use the similarity measure from the inner product of the score vector of the user and that of the advertisement. It is possible  that in a cascade of two users $\{u_1, u_2\}$,
both users like the advertisement because their inner products are large,
but
user $u_1$ likes one topic in that advertisement and user $u_2$ likes another different topic in that advertisement.
Such a cascade cannot be caught by using the two-step approach in TwitterRank \cite{weng2010twitterrank}.
In view of this, it might be better to use a single-step approach for computing advertisement-specific influence centralities. 
As the influence centralities in the previous section, we propose using opinion dynamics through a path. 
For a path $r$ that traverses a sequence of nodes $\{u_1, u_2,\ldots, u_{k-1}, u_k\}$ in the attributed network, we  define the following path measure
\beq{topic3333}
f(r)=\min_{u_i\in r}[z \cdot h_V(u_i)],
\eeq
where
$$z \cdot h_V(u_i)=\sum_{i=1}^p z_i \cdot h_{V,i}(u_i)$$
is the inner product of $z$ and $h_V(u)$,
If we use the random walk with path length 1 or 2 to sample a path $r$ in $G(V,E)$ as in \req{path2222}, then
the sampling distribution for the attributed network with the path measure in \req{topic3333} can be written as follows:
\bearno
pp(r)=\left\{\begin{array}{ll}
                 C \cdot e^{\theta \min[z\cdot h_V(u_1) , z \cdot h_V(u_2)]} \cdot {{\beta_1} \over {2m}}a_{u_1,u_2}, & \\ \quad\quad \mbox{if}\; r=\{u_1,u_2\}, & \\
                 \\
                C \cdot e^{\theta \cdot \min[z\cdot h_V(u_1) , z \cdot h_V(u_2),z \cdot h_V(u_3)]} \cdot {{\beta_2} \over {2m}} \frac{a_{u_1, u_2} a_{u_2,u_3}}{k_{u_2}}, & \\ \quad\quad \mbox{if}\; r=\{u_1,u_2,u_3\}. &
                \end{array} \right.
\eearno
\beq{topic4444}
\eeq
Moreover, we have the following bivariate distribution
\bear{topic5555}
&&p_{U,W}(u,w)=C \Big [ e^{\theta \min[z\cdot h_V(u) , z \cdot h_V(w)]} \cdot {{\beta_1} \over {2m}}a_{u,w} \nonumber \\
&&+\sum_{u_2 \in V} e^{\theta \cdot \min[z\cdot h_V(u) , z \cdot h_V(u_2),z \cdot h_V(w)]} \cdot {{\beta_2} \over {2m}} \frac{a_{u, u_2} a_{u_2,w}}{k_{u_2}}\Big ],\nonumber\\
\eear
where $C$ is the normalization constant.
The marginal distribution of the bivariate distribution, denoted by $P_U(u)$, is the {\em advertisement-specific influence centrality} of node $u$ for advertisement $z$ (with respect to the temperature $\theta$).

\section{Experiment}
\subsection{Experiment for influence centrality}
In this section, we evaluate our model on the real-world network proposed in \cite{adamic2005political} which is a directed network of hyperlinks between weblogs of US politics recorded by Adamic and Glance. The data on political leaning come from blog directories as indicated. Some blogs were labeled manually, based on incoming and outgoing links and posts around the time of the United States presidential election of 2004. Basically, there are 1,490 nodes and 19,090 edges in this dataset. For our purposes, we would like to examine our algorithm in a undirected signed network. Therefore, we choose the nodes from different community and adding 4,000 negative edges to this network. As a reason that every node should be treated fairly, we randomly pick up the node in the different community instead of choosing those with a higher degree. Then let the the adjacency matrix $A=A+A^T$ because we want to get the undirected network. Before we rank these nodes, we have to delete the nodes with degree smaller than 7, that is, their ranks are not important. Furthermore, we also remove the edges which are self-loop and the edges between two nodes if they are connected by more than one edges. As a result, we obtain the directed signed network containing 863 nodes and 16,650 edges, including 15,225 positive edges and  1,425 negative edges.

\begin{figure*}[tb]
	\centering
    \begin{tabular}{p{0.32\textwidth}p{0.32\textwidth}p{0.32\textwidth}}
      \includegraphics[width=0.32\textwidth]{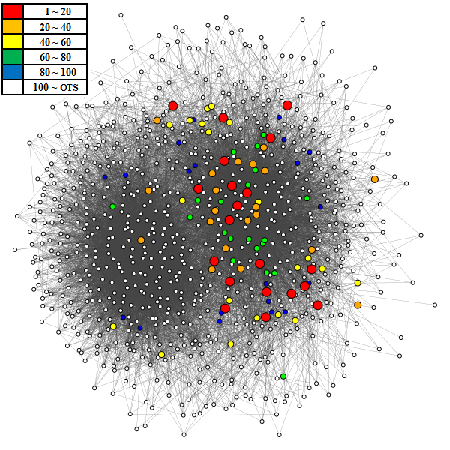} &
      \includegraphics[width=0.32\textwidth]{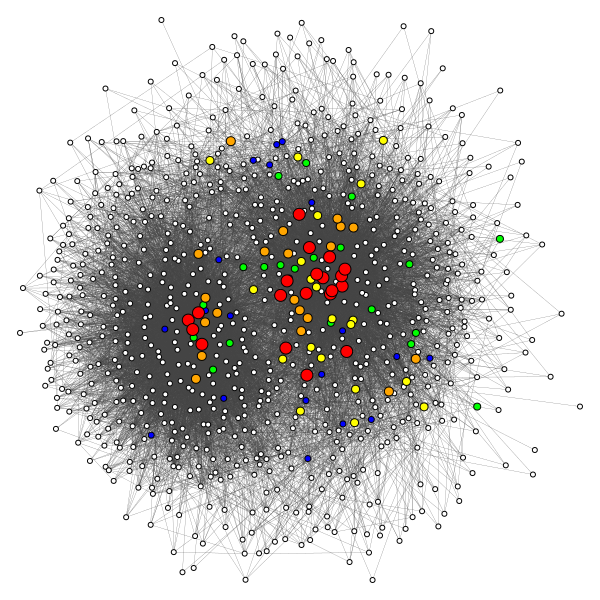} &
      \includegraphics[width=0.32\textwidth]{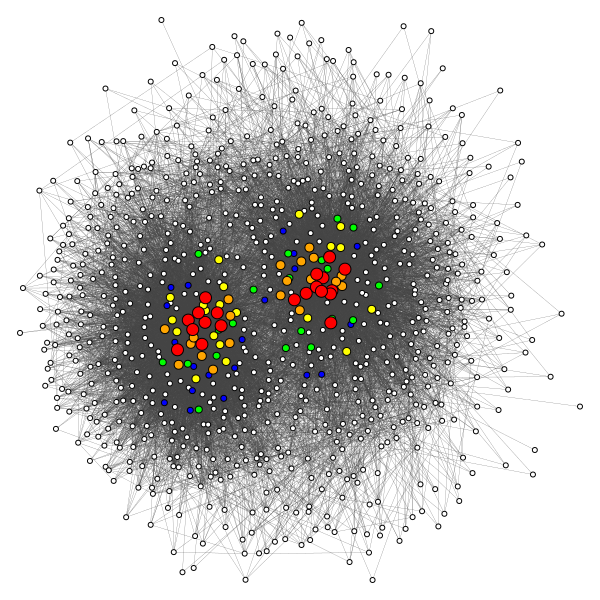} \\
      (a) $\gamma$ = -0.9, $\theta$=-2.6566 & (b) $\gamma$ =-0.5, $\theta$=-1.7337 & (c) $\gamma$ =0, $\theta$=-1.1844 \\
      \includegraphics[width=0.32\textwidth]{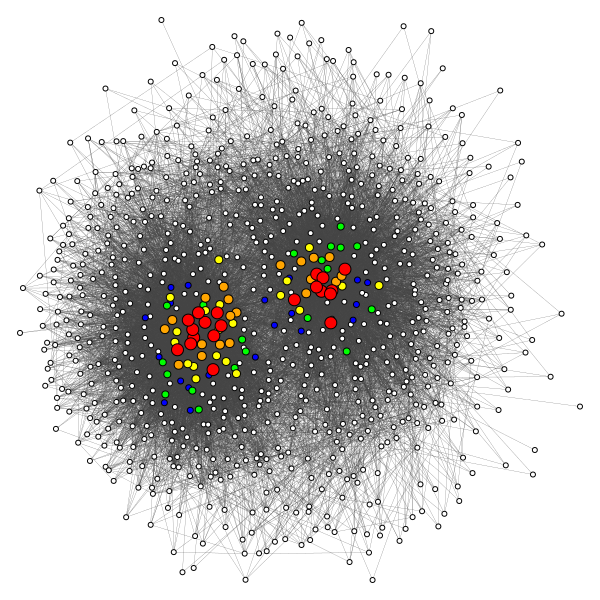} &
      \includegraphics[width=0.32\textwidth]{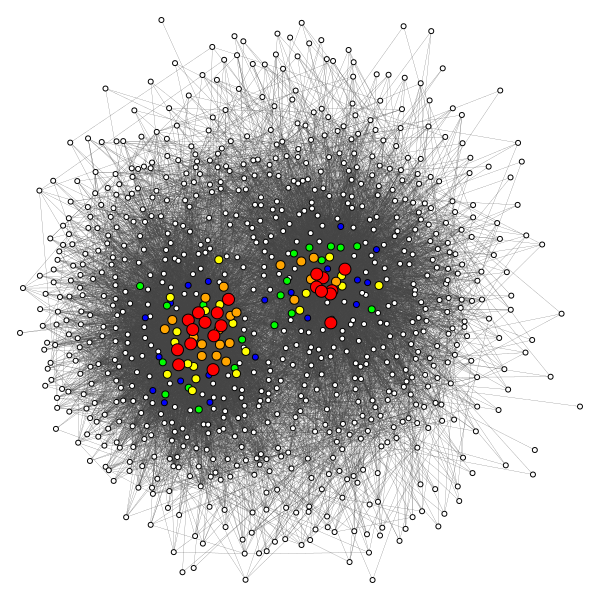} &
      \includegraphics[width=0.32\textwidth]{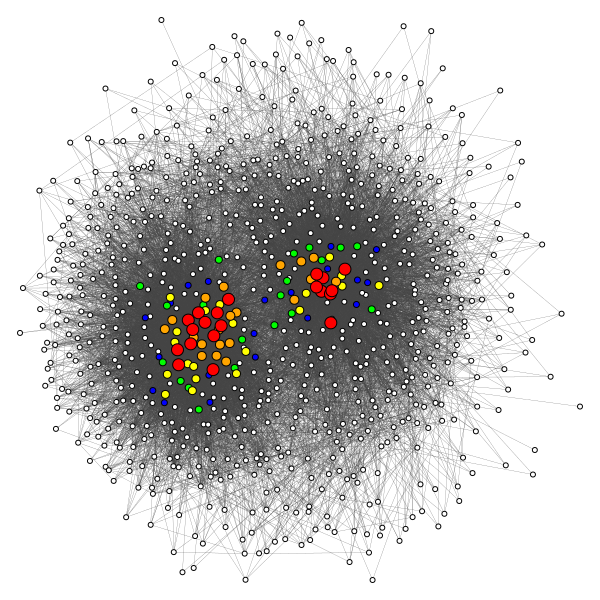}\\
      (d) $\gamma$ =0.5, $\theta$=-0.6351 & (e) $\gamma$ =0.9, $\theta$=0.2878 & (f) $\gamma$ =0.99, $\theta$=1.4623 \\ \\
    \end{tabular}
    \caption{The ranking result of influence centralities by $\beta_2$=0 with different theta}
    \label{fig:colornode}
\end{figure*}

\begin{table}
{\footnotesize
\begin{center}
\caption{The Jaccard index results of influence centralities for various $\gamma$ settings by $\beta_2$=0}
\label{tab:Jac1}
\begin{tabular}{|c|c|c|c|}
\hline
\multirow{2}{*}{}                                                                          & \multirow{2}{*}{\begin{tabular}[c]{@{}c@{}}ranking by\\ positive edge\end{tabular}} & \multirow{2}{*}{\begin{tabular}[c]{@{}c@{}}ranking by\\ negative edge\end{tabular}} & \multirow{2}{*}{\begin{tabular}[c]{@{}c@{}}ranking by\\ total edge\end{tabular}} \\
                                                                                           &                                                                                     &                                                                                     &                                                                                  \\ \hline
\multirow{2}{*}{\begin{tabular}[c]{@{}c@{}}$\gamma$=-0.99\\ $\theta$=-3.8310\end{tabular}} & \multirow{2}{*}{0.0417}                                                              & \multirow{2}{*}{0.7544}                                                             & \multirow{2}{*}{0.0417}                                                          \\
                                                                                           &                                                                                     &                                                                                     &                                                                                  \\ \hline
\multirow{2}{*}{\begin{tabular}[c]{@{}c@{}}$\gamma$=-0.9 \\ $\theta$=-2.6566\end{tabular}}  & \multirow{2}{*}{0.0753}                                                              & \multirow{2}{*}{0.6667}                                                             & \multirow{2}{*}{0.0753}                                                          \\
                                                                                           &                                                                                     &                                                                                     &                                                                                  \\ \hline
\multirow{2}{*}{\begin{tabular}[c]{@{}c@{}}$\gamma$=-0.5\\ $\theta$=-1.7337\end{tabular}}  & \multirow{2}{*}{0.2658}                                                              & \multirow{2}{*}{0.3514}                                                             & \multirow{2}{*}{0.2658}                                                          \\
                                                                                           &                                                                                     &                                                                                     &                                                                                  \\ \hline
\multirow{2}{*}{\begin{tabular}[c]{@{}c@{}}$\gamma$=0\\ $\theta$=-1.1844\end{tabular}}     & \multirow{2}{*}{0.7241}                                                             & \multirow{2}{*}{0.0989}                                                               & \multirow{2}{*}{0.7241}                                                          \\
                                                                                           &                                                                                     &                                                                                     &                                                                                  \\ \hline
\multirow{2}{*}{\begin{tabular}[c]{@{}c@{}}$\gamma$=0.5\\ $\theta$=-0.6351\end{tabular}}   & \multirow{2}{*}{0.8868}                                                             & \multirow{2}{*}{0.0638}                                                              & \multirow{2}{*}{0.8868}                                                          \\
                                                                                           &                                                                                     &                                                                                     &                                                                                  \\ \hline
\multirow{2}{*}{\begin{tabular}[c]{@{}c@{}}$\gamma$=0.9\\ $\theta$=0.2878\end{tabular}}    & \multirow{2}{*}{1}                                                                  & \multirow{2}{*}{0.0417}                                                             & \multirow{2}{*}{0.9608}                                                          \\
                                                                                           &                                                                                     &                                                                                     &                                                                                  \\ \hline
\multirow{2}{*}{\begin{tabular}[c]{@{}c@{}}$\gamma$=0.99\\ $\theta$=1.4323\end{tabular}}   & \multirow{2}{*}{1}                                                                  & \multirow{2}{*}{0.0417}                                                             & \multirow{2}{*}{0.9608}                                                          \\
                                                                                           &                                                                                     &                                                                                     &                                                                                  \\ \hline

\end{tabular}
\end{center}}
\end{table}

\subsubsection{Experiment by $\beta_2$=0}

In \rfig{colornode}, we show the distribution of ranking by $\beta_2$=0. Initially, the top 100 nodes are uniformly distributed on the right side on (a). However, when the value of gamma increases, the top 100 nodes gradually gather in the middle of two community where nodes have more edges than those on the edge of the graph. Also, these top 100 nodes are divided into two clusters and for the nodes in each cluster, they are closely compact with each other. From so, the key persons for these two clusters can be found at ease. On the other hand, we can get the information of which nodes are affecting others in a negative way while gamma is lower. That is to say; these nodes usually have a much higher negative degree compared to others. We can view the nodes with a higher negative degree as people who are sent to attack the opponent team, which we call a hitman in this paper. Since these graphs only provide the fact that the top 100 nodes would gather in the middle of the graph if gamma is close to 1, we use Jaccard index for more detail explanation. The Jaccard index is a statistic used for comparing the similarity and diversity of sample sets. It can be denoted as 
\beq{Jaccard}
J(S_1,S_2)=\frac{|S_1 \cap S_2|}{|S_1 \cup S_2|},
\eeq
where $S_1$ and $S_2$ are the sample sets.

\begin{figure}[h]
\centering
\includegraphics[width=0.7\textwidth]{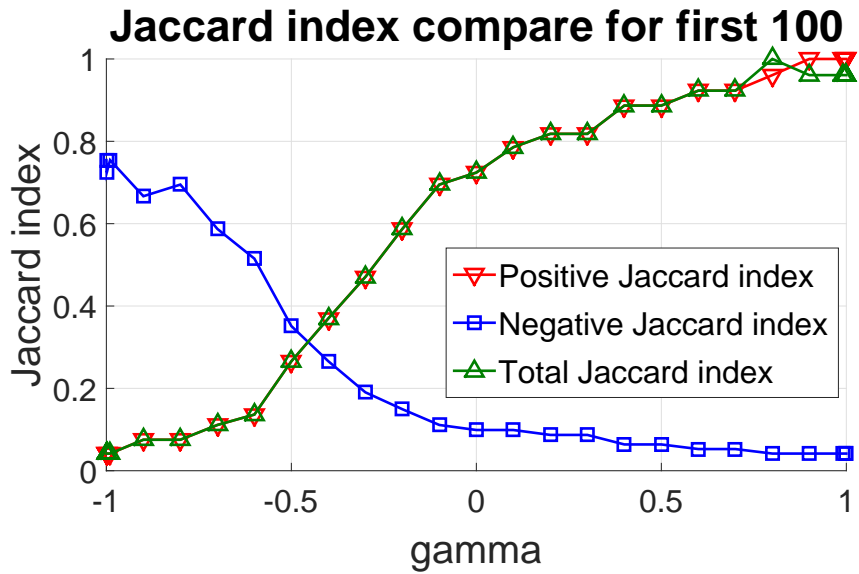}
\caption{The Jaccard index compare of influence centralities by $\beta_2=0$}
\label{fig:jar}
\end{figure}

In \rtab{Jac1}, we use three kinds of Jaccard index as metrics for comparing the similarity between two sample sets. The negative Jaccard index, the similarity between the set ranked by negative edges and the set ranked by $\beta_2=0$. Similarly, the positive (total) Jaccard index is the similarity between the set ranked by positive (total) edges and the set ranked by $\beta_2=0$. It is easy for us to conclude that the increment of gamma has a positive impact on raising theta. Also, the change of theta causes the change of all three Jaccard indexes. More specifically, the negative Jaccard index decreases and the positive Jaccard index increases along with the increment of theta. Interestingly, the total Jaccard index has a larger value when theta is closer to zero. In order to look into the relationship between these three parameters, \rfig{jar} is plotted.

\begin{figure*}[tb]
	\centering
    \begin{tabular}{p{0.32\textwidth}p{0.32\textwidth}p{0.32\textwidth}}
      \includegraphics[width=0.32\textwidth]{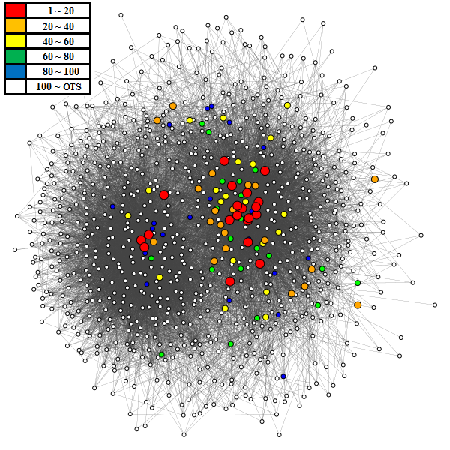} &
      \includegraphics[width=0.32\textwidth]{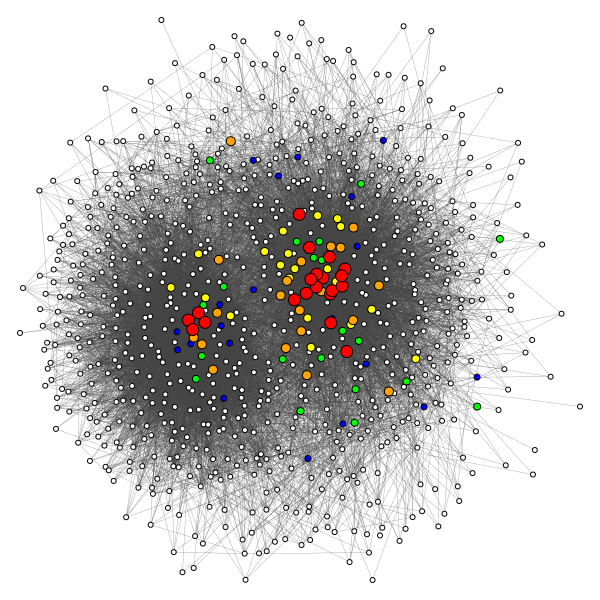} &
      \includegraphics[width=0.32\textwidth]{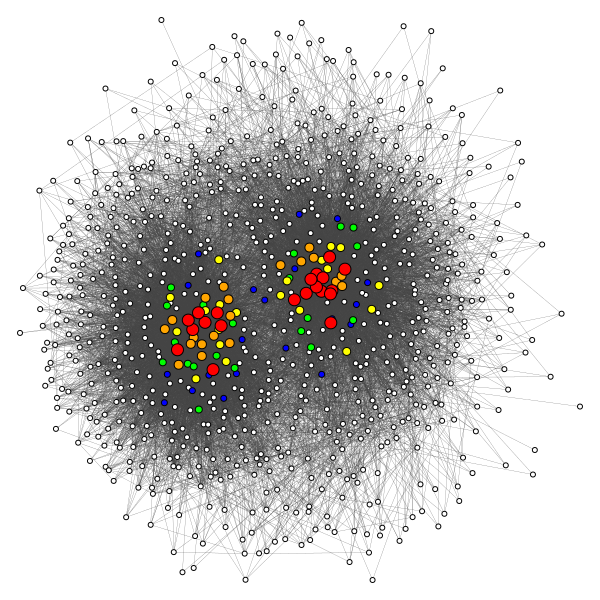} \\
      (a) $\theta$ =-3 & (b) $\theta$ =-2 & (c) $\theta$ =-1 \\
      \includegraphics[width=0.32\textwidth]{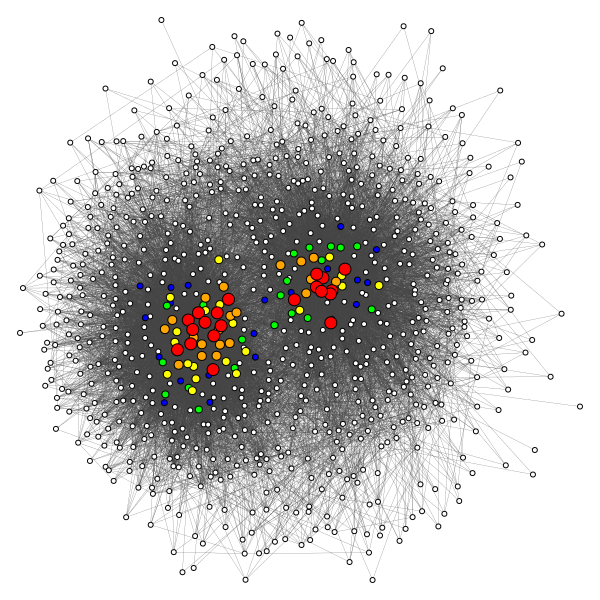} &
      \includegraphics[width=0.32\textwidth]{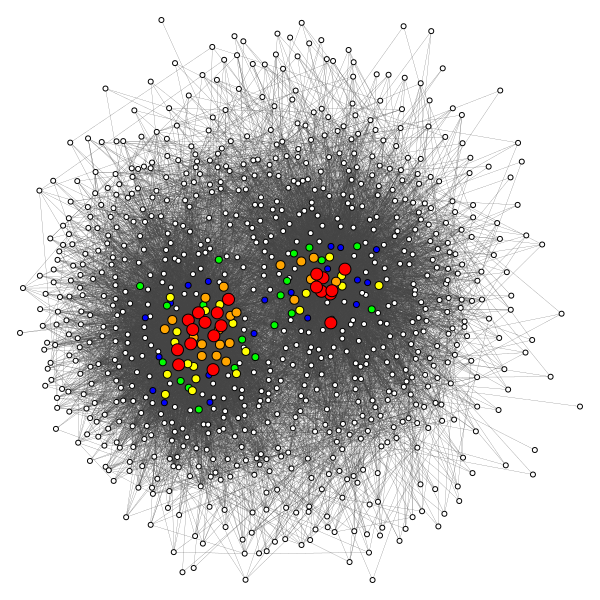} &
      \includegraphics[width=0.32\textwidth]{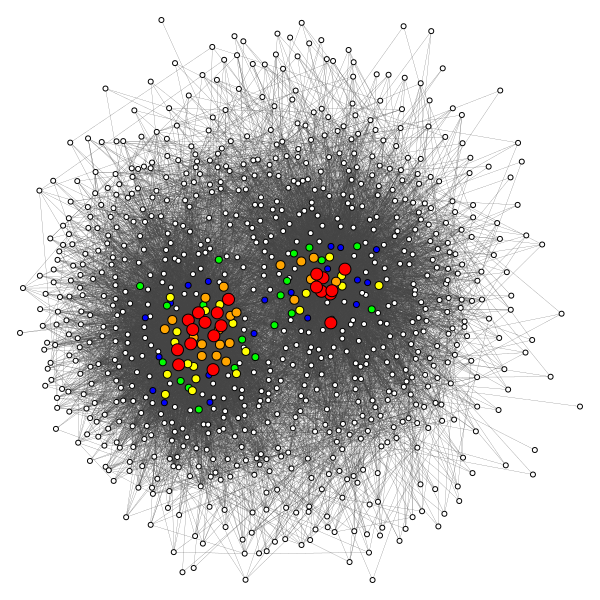}\\
      (d) $\theta$ =0 & (e) $\theta$ =1 & (f) $\theta$ =2 \\ \\
    \end{tabular}
    \caption{The ranking result of influence centralities by $\beta_1=0.7$ and $\beta_2=0.3$}
    \label{fig:colornode2}
\end{figure*}

In \rfig{jar}, we can make a conclusion that the positive Jaccard index and the negative Jaccard index have positive and negative relation with gamma respectively.  For the total Jaccard index, it has its maximum when gamma is close to 0.8. This implies the fact that the closer to 0.8 the gamma is, the closer to zero the theta is. As a result, we can use this fact as a pioneering method to define the centrality for a signed network. That is to say, the centrality can be defined by the adjustment of temperature (theta). By using opinion dynamic, there are three different centralities able to be defined. First, the positive influence centrality, representing those people who give you positive influences (e.g., your friends), can be found by increasing the temperature (theta). On the contrary, the negative influence centrality stands for those people giving you negative influences like your enemy. We figure it out by lowering down the temperature (theta). Last, the total influence centrality is the representation for the famous people among all. For these famous people, the thing we only care about is how many people know him instead of how many people love him or hate him. Thus, we make the temperature (theta) close to zero to find the total influence centrality.

\begin{table}
{\footnotesize
\begin{center}
\caption{The Jaccard index results of influence centralities for various $\theta$ settings by $\beta_1$=0.7 and $\beta_2$=0.3.}
\label{tab:Jac2}
\begin{tabular}{|c|c|c|c|}
\hline
\multirow{2}{*}{} & \multirow{2}{*}{\begin{tabular}[c]{@{}c@{}}ranking by\\ positive edge\end{tabular}} & \multirow{2}{*}{\begin{tabular}[c]{@{}c@{}}ranking by\\ negative edge\end{tabular}} & \multirow{2}{*}{\begin{tabular}[c]{@{}c@{}}ranking by\\ total edge\end{tabular}} \\
                  &                                                                                     &                                                                                     &                                                                                  \\ \hline
$\theta$=-2       & 0.3423                                                                              & 0.3423                                                                              & 0.3605 \\ \hline
$\theta$=-1.5     & 0.4925                                                                              & 0.1976                                                                              & 0.5152 \\ \hline
$\theta$=-1       & 0.6949                                                                              & 0.1299                                                                              & 0.7241                                                                           \\ \hline
$\theta$=-0.5     & 0.8868                                                                              & 0.0753                                                                              & 0.9231                                                                           \\ \hline
$\theta$=0        & 0.9608                                                                              & 0.0638                                                                              & 1                                                                                \\ \hline
$\theta$=0.5      & 0.9802                                                                              & 0.0638                                                                              & 0.9802                                                                           \\ \hline
$\theta$=1        & 1                                                                                   & 0.0582                                                                              & 0.9608                                                                           \\ \hline
$\theta$=1.5      & 1                                                                                   & 0.0582                                                                              & 0.9608                                                                           \\ \hline
$\theta$=2        & 1                                                                                   & 0.0582                                                                              & 0.9608                                                                           \\ \hline
\end{tabular}
\end{center}}
\end{table}

\begin{figure}[h]
\centering
\includegraphics[width=0.7\textwidth]{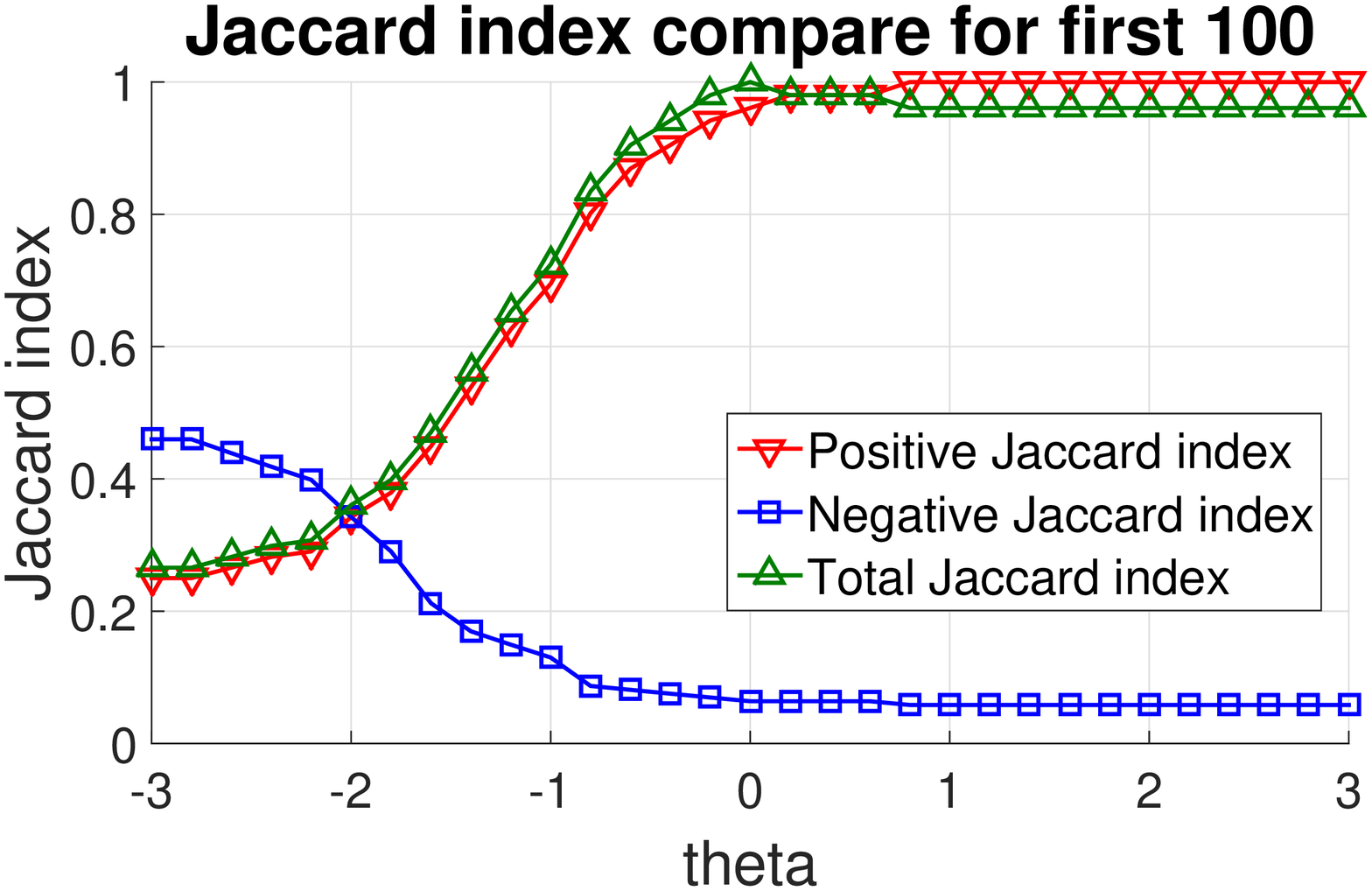}
\caption{The Jaccard index compare of influence centralities by $\beta_1$=0.7 and $\beta_2$=0.3.}
\label{fig:jar2}
\end{figure}

\begin{figure*}[tb]
	\centering
    \begin{tabular}{p{0.32\textwidth}p{0.32\textwidth}p{0.32\textwidth}}
      \includegraphics[width=0.32\textwidth]{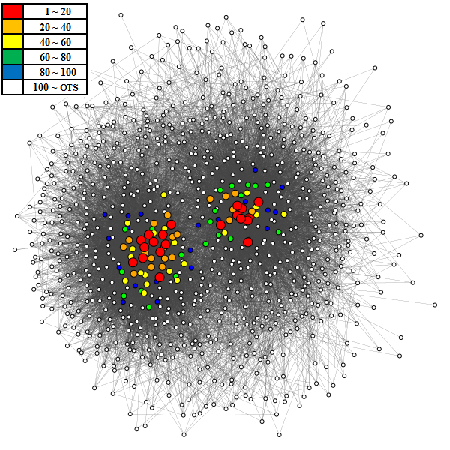} &
      \includegraphics[width=0.32\textwidth]{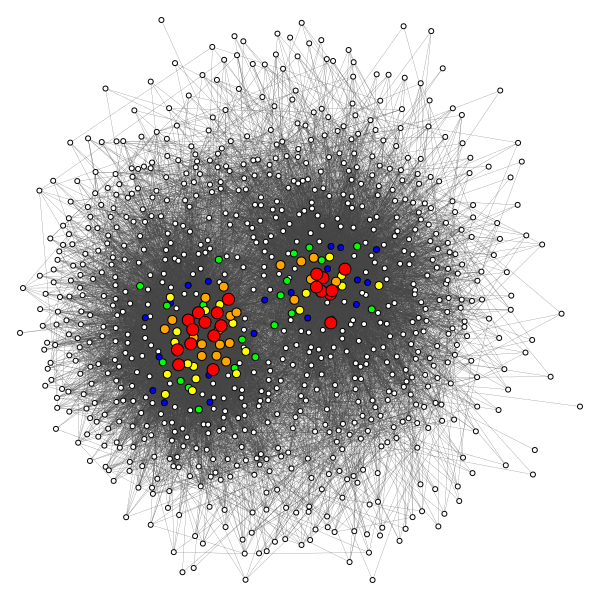} &
      \includegraphics[width=0.32\textwidth]{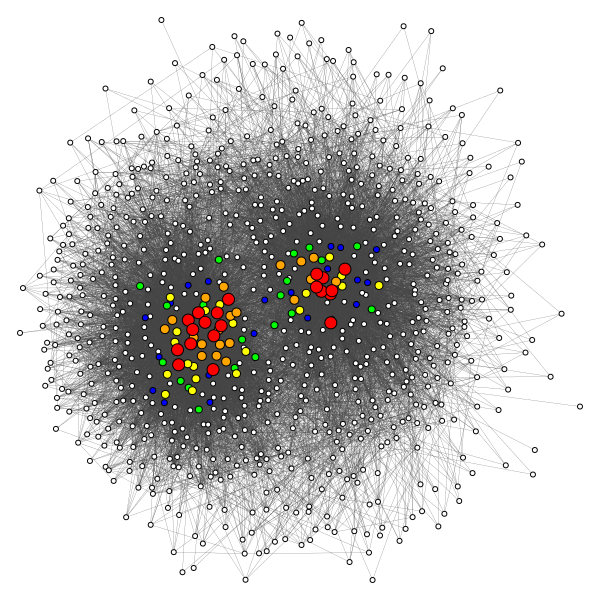} \\
      (a) $\theta$ =0 & (b) $\theta$ =1 & (c) $\theta$ =2 \\
    \end{tabular}
    \caption{The ranking result of trust centralities by $\beta_1$=0.7 and $\beta_2$=0.3 with different theta}
    \label{fig:trusttendency}
\end{figure*}

\subsubsection{Experiment by $\beta_1=0.7$ and $\beta_2=0.3$}

In \rfig{colornode2}, we show the distribution of ranking by $\beta_1=0.7$ and $\beta_2=0.3$. Same as \rfig{colornode}, theta has the same effect for gathering node in the middle of graphs. In addition, we can observe that the hitmen by setting $\beta_1=0.7$ and $\beta_2=0.3$ are slightly different from ranking by $\beta_2=0$. The reason is that when we have the information from our neighbors' neighbor, some messages that our neighbors are unwilling to tell may be revealed. Therefore, these hidden hitmen will emerge due to the exposure of these secret messages. More details are provided in \rtab{Jac2}.

In \rtab{Jac2}, the values for the three Jaccard indexes we proposed are listed. As the theta increases, the positive Jaccard index does increase as a very similar way as listed in \rtab{Jac1}. However, the positive Jaccard index in \rtab{Jac1} is one when $\theta$ is 0.2878 and the positive Jaccard index is one when $\theta$ is one. We can find if we consider the path which length is two, the positive Jaccard index converges slowly. It is because we have to consider more information than only consider the path which length is one. The obvious difference is that the negative Jaccard index values are much lower compared to the result that only consider the path which length is one. Due to the fact that when considering not just the neighbors of a node but both its neighbors and its neighbors' neighbors, we would get more information and thus find a more global ranking. The negative Jaccard index and the total Jaccard index have very similar changes of behavior as the positive Jaccard index when considering the path which length is two. To study more about the difference between these two path sampling, \rfig{jar2} provides more graphical details.

For the negative Jaccard index, it is relatively lower compared to the result of considering the path which length is one in \rfig{jar2}. Since considering the path which length is two would consider two edges rather than one, it means that there will be four circumstances to consist a two-step edge pair including negative/negative, negative/positive, positive/negative and positive/positive. We have to consider the positive/negative and negative/positive edge pairs to figure out the negative Jaccard index. Therefore, the lower negative Jaccard index has been generated. In the same way, we have to add the positive/positive and negative/negative edge pairs into our consideration for the calculation of the positive Jaccard index.  We also look forward to a relatively lower positive Jaccard index when $\theta$ is not too large same as the negative Jaccard index does. However, there are only 8.5\% of negative edges in this network such that the influence of negative edges can be ignored. As a result, the positive Jaccard index acts in the same way as it does in considering the path which length is only one when theta grows larger.

In opinion dynamic, we say that the spread of the information is a way to define centrality. Considering the path which length is only one provides a way to find the influence centrality for a node to its neighbors. Moreover, if we want to find the influence centrality for a node to those nodes which can not be reached in random walk only one step, we have to apply random walk more step to figure this out.

\begin{figure}[h]
\centering
\includegraphics[width=0.7\textwidth]{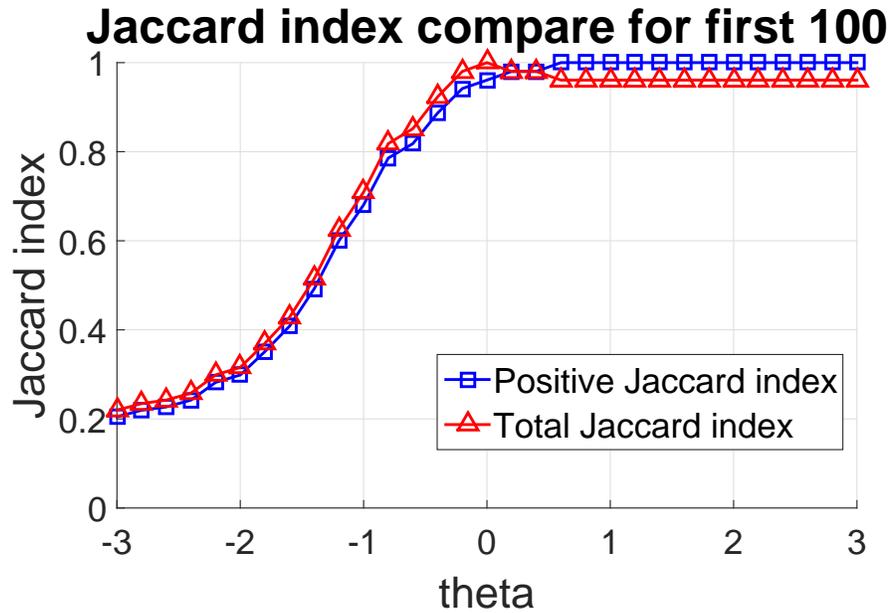}
\caption{The Jaccard index compare of trust centralities.}
\label{fig:Trustjar}
\end{figure}

\subsection{Experiment for trust centrality}
We use the same data as used in the experiment for influence centrality. In trust centrality, we have to consider the path which length is two because if only considering the path which length is one the result is as same as the influence centrality considering only the path which length is one. In \rfig{trusttendency}, no matter how we adjust theta, figure (a), (b) and (c) look similar. It is because the higher ranking node has more positive edges in trust centrality, and we have to consider positive/positive edges in two-step approach. Then in this dataset, the positive edges accounted for 91.5\%. When theta is closer to zero, the rank is more similar to rank by total edges. When theta is larger, the rank is more similar to rank by positive edges. It is different from \rfig{jar2}.
 
In \rfig{jar2}, the line of positive Jaccard index slowly close to one when theta is larger. It is because it has to consider the negative/negative and positive/positive edges. In \rfig{Trustjar}, we only have to consider positive/positive edges, so the line is equal to one when theta is larger. It is as same as the line in \rfig{jar}. With the increase of theta, we can find who is worthy of our trust. It is our definition of trust centrality.

\begin{figure*}[tb]
	\centering
    \begin{tabular}{p{0.32\textwidth}p{0.32\textwidth}p{0.32\textwidth}}
      \includegraphics[width=0.32\textwidth]{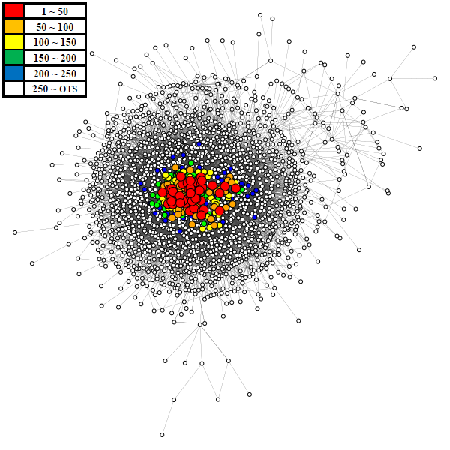} &
      \includegraphics[width=0.32\textwidth]{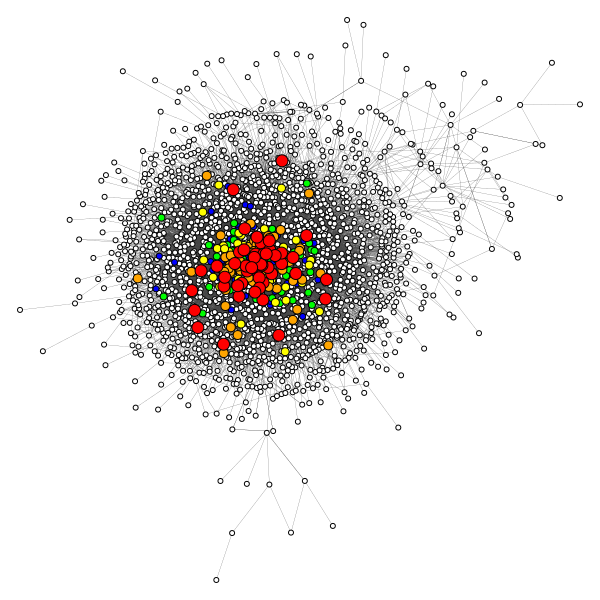} &
      \includegraphics[width=0.32\textwidth]{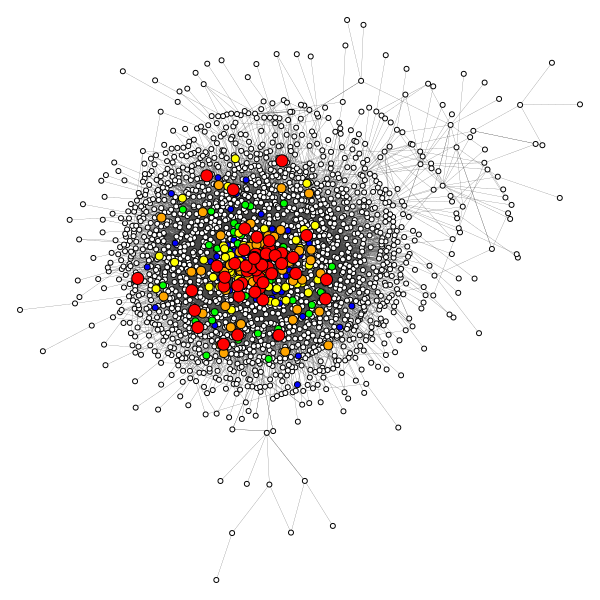} \\
      (a) $\theta$ =0 & (b) $\theta$ =0.05 & (c) $\theta$ =0.1 \\
    \end{tabular}
    \caption{The ranking result of advertisement-specific influence centralities with different theta}
    \label{fig:topictendency}
\end{figure*}

\subsection{Experiment for advertisement-specific influence centrality}

We evaluate our model on MemeTracker dataset \cite{leskovec2008memetracker}, which tracks the quotes and phrases that frequently appear over time across mass media and personal blogs. Each cascade is composed of a "meme' (the phrase being mentioned), and the timestamps records of sites which have mentioned that phrase. To get advertisement information from these memes, we leverage Carrot2 \cite{osinski2005carrot2}, an open source clustering engine, with STC clustering algorithm to classify memes into fifteen topics including People, Going, Know, Years, Way, United States, States, Life, Believe, Lot, Love, America, Country, Barack Obama and Obama. For the data-independency purpose, we delete United States and Barack Obama since they are subsets of topic States and Obama. Therefore, we obtain a dataset with thirteen regularly-used phrases in it. Before we rank these nodes, we have to delete the nodes with degree smaller than 7 and self-loop edges, that is, their ranks are not important. As a result, we obtain the attributed network containing 2082 nodes and 16,503 edges.

We choose the phrase "Going" as an advertisement and input it into the existing attributed network that has node attribute. Since the ranking results by the sampling with path length 1 and the sampling with path length 2 are almost the same, we simply use the rank of sampling with path length 1 to do performance evaluation. Then, treating the advertisement as a sampling bias to find the pass measure and plot it as \rfig{topictendency} varying with $\theta$. As so, it is obvious to see that the red nodes with higher rank gradually gather in the middle of the graph while $\theta$ slowly decrease down to zero. Then, we use Jaccard index for more detail explanation.

\begin{figure}[h]
\centering
\includegraphics[width=0.7\textwidth]{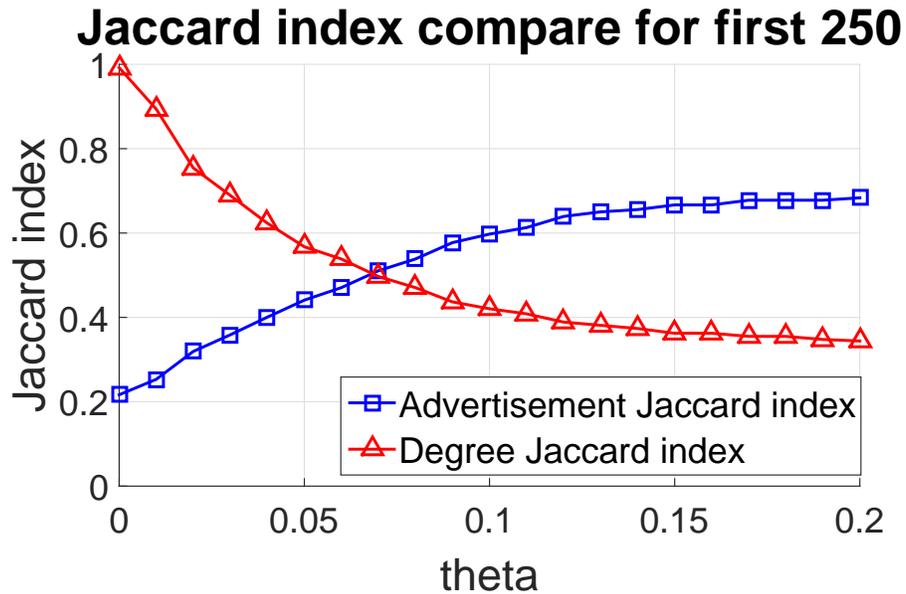}
\caption{The Jaccard index compare of advertisement-specific influence centralities.}
\label{fig:Topicjar}
\end{figure}

In this paper, we simply call the Jaccard index between the rank of degree and the rank generated by setting $\beta_2=0$ as Degree Jaccard index. Also, we call the Jaccard index between the rank of the regularly-used phrases and rank generated by setting the $\beta_2=0$ as Advertisement Jaccard index. In \rfig{Topicjar}, we show the relationship between $\theta$ and the two Jaccard index respectively. Degree Jaccard index has an escalating trend while $\theta$ is close to zero. On the other hand, Advertisement Jaccard index raises up as $\theta$ grows large, we claim it as advertisement-specific influence centrality. Then, the smaller the $\theta$ is, the more similar the ranking is to the rank of degree. On the contrary, if the $\theta$ increases, the rank would be more relevant to the rank of advertisement.

\begin{figure*}[tb]
	\centering
    \begin{tabular}{p{0.32\textwidth}p{0.32\textwidth}p{0.32\textwidth}}
      \includegraphics[width=0.32\textwidth]{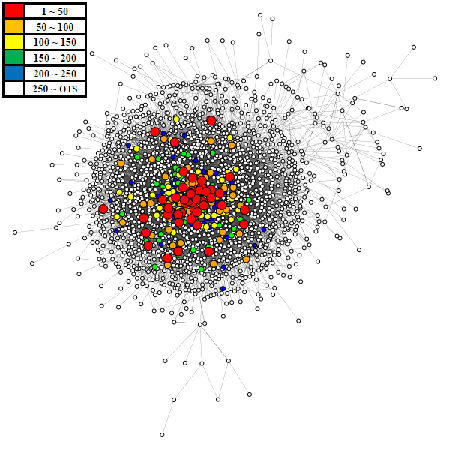} &
      \includegraphics[width=0.32\textwidth]{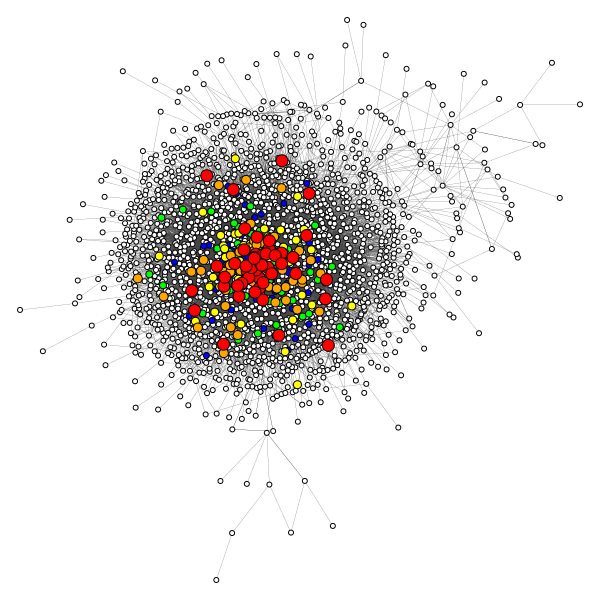} &
      \includegraphics[width=0.32\textwidth]{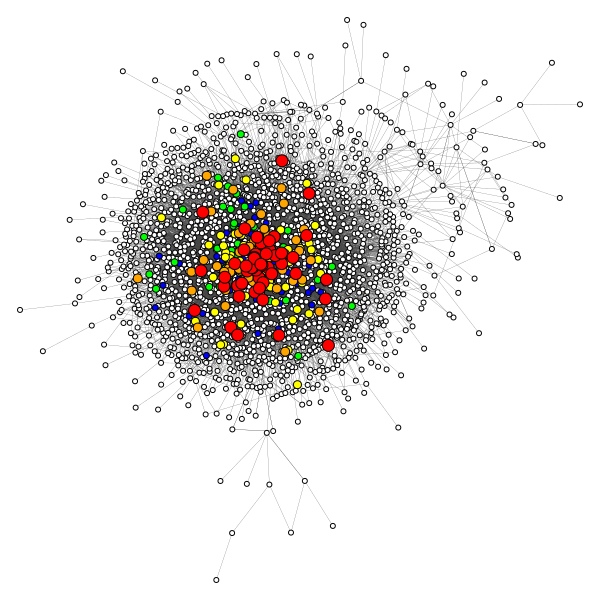} \\
      (a) Going & (b) Know & (c) People \\
      \includegraphics[width=0.32\textwidth]{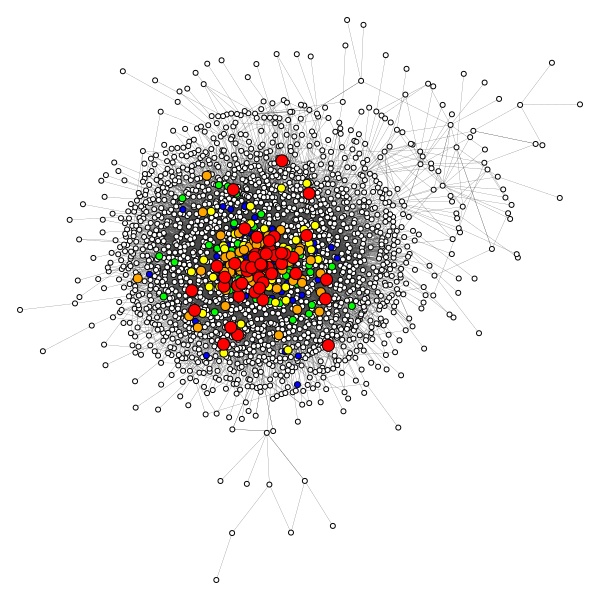} &
      \includegraphics[width=0.32\textwidth]{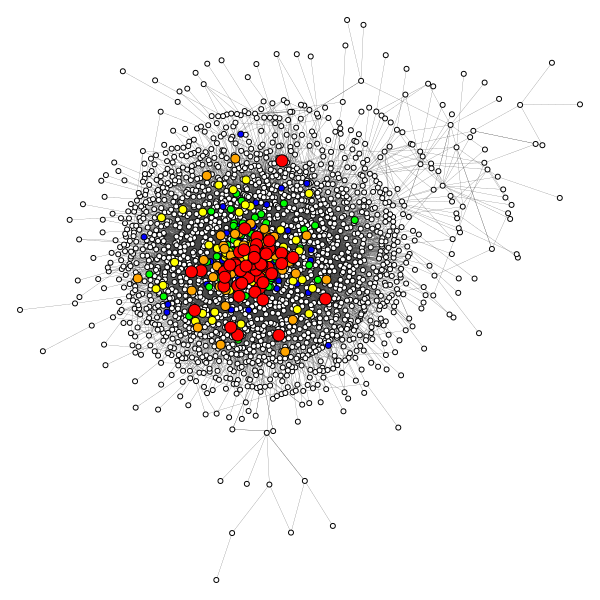} &
      \includegraphics[width=0.32\textwidth]{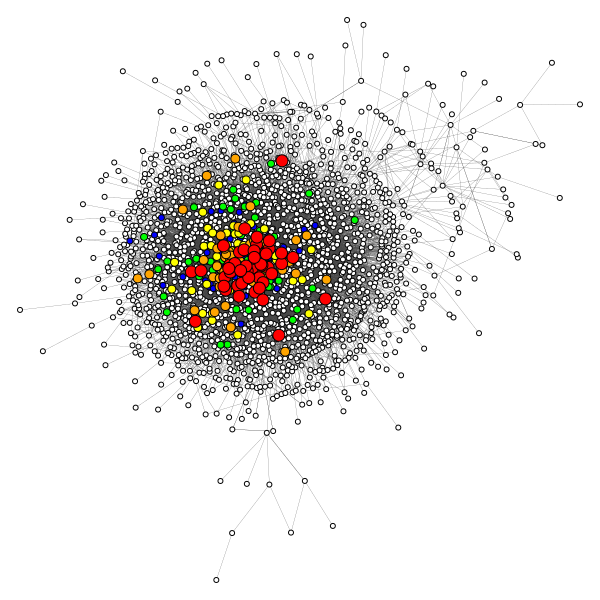} \\
      (a) Years & (b) America & (c) Obama \\
    \end{tabular}
    \caption{The ranking result of advertisement-specific influence centralities with different advertisement}
    \label{fig:differenttopic}
\end{figure*}

For experimental settings, we set $\theta$ as 0.2 because the top 250 nodes do not change by increasing the $\theta$ over than 0.2. Then we choose the top six most frequently-used phrases as test data, including Going, Know, People, Years, America, and Obama. We apply each phrase to our algorithm one by one and find that the top rank nodes are not in complete accord. \rfig{differenttopic} leads to the fact that the top rank nodes may not be the same between any two topics.

\begin{figure*}[tb]
	\centering
    \begin{tabular}{p{0.45\textwidth}p{0.45\textwidth}}
      \includegraphics[width=0.45\textwidth]{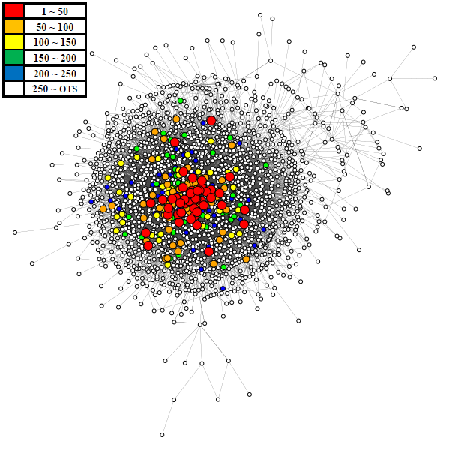} &
      \includegraphics[width=0.45\textwidth]{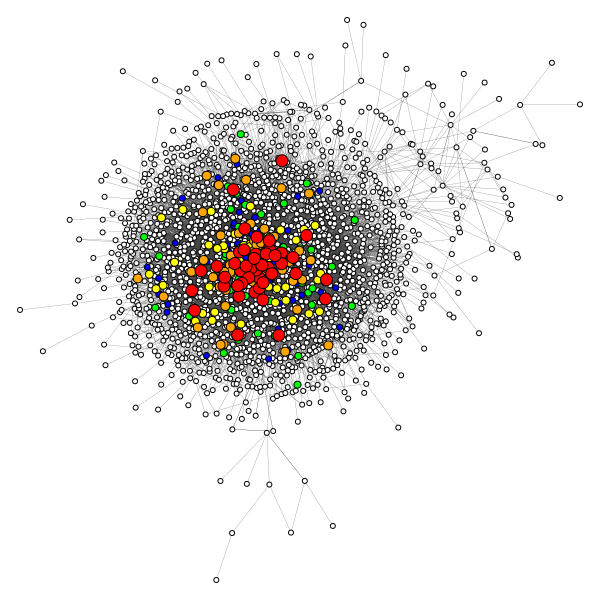} \\
      (a) Going and Obama  & (b) Know and America \\
    \end{tabular}
    \caption{The ranking result of advertisement-specific influence centralities with combining two advertisement}
    \label{fig:twotopic}
\end{figure*}
Besides, we combine advertisements from different aspects and do the performance evaluation to find the top rank node of these advertisements. In \rfig{twotopic}, figure (a) is combined by two advertisements that is "Going" and "Obama," and figure (b) is combined by the advertisement "Know" and the advertisement "America". The result is not necessarily the same compared to the result of each advertisement we used to combine in.

As a result, if there is an advertisement containing all sort of contents and we would like to know who is the most influential person to it in the existing network, we can get the answer through this experiment. Briefly speaking, we pick up those important phrases in the contents manually and apply them to our algorithms. Thus, those nodes which are truly beneficial to the promotion for the advertisement can be found out. This may be very different to the fact that people usually think the higher degree a node has the more value the node can create for an advertisement.

\section{Conclusion}


In this paper, we extend our previous work in \cite{chang2016probabilistic} to {\em attributed networks}. Our approach is to extend the bivariate distributions in an unsigned network to a signed network by using the exponential change of measure. Hence, the centrality is also being separated into three different terms of centrality which are the positive influence centrality, negative influence centrality, and the total influence centrality. Besides, the experimental results also help us learn the relationship between temperature (theta) and the three centralities. We also would like to extend this definition to community detection and not just using it as centrality.

\bibliographystyle{IEEEtran}
\bibliography{CHChang}

\end{document}